\documentclass[11pt,letterpaper]{article}
\usepackage{systeme} 
\usepackage[utf8]{inputenc}
\usepackage{multirow} 
\usepackage{booktabs} 
\usepackage{array} 
\usepackage[english]{babel}
\usepackage{amsmath}
\usepackage{amsfonts}
\usepackage{amssymb}
\usepackage{graphicx}
\usepackage[small,bf]{caption} 
\usepackage[left=3cm,right=3cm,top=2cm,bottom=3cm]{geometry}
\usepackage{framed} 
\usepackage{color} 
\usepackage{wrapfig}\definecolor{shadecolor}{RGB}{230,230,230} 

\author{Jorge Pinochet}
\title{\textbf{Advanced physics transposition (APT): A new name for a not-so-new field of research}}

\begin{document}

\author{Jorge Pinochet$^{*}$\\ \\
 \small{$^{*}$\textit{Facultad de Ciencias Básicas, Departamento de Física. }}\\
  \small{\textit{Centro de Investigación en Educación (CIE-UMCE),}}\\
 \small{\textit{Núcleo Pensamiento Computacional y Educación para el Desarrollo Sostenible (NuCES).}}\\
 \small{\textit{Universidad Metropolitana de Ciencias de la Educación,}}\\
 \small{\textit{Av. José Pedro Alessandri 774, Ñuñoa, Santiago, Chile.}}\\
 \small{e-mail: jorge.pinochet@umce.cl}\\}

\date{}
\maketitle

\begin{center}\rule{0.9\textwidth}{0.1mm} \end{center}
\begin{abstract}
\noindent Research into educational physics is a field of study that has undergone sustained growth in recent decades. Among the topics addressed in educational physics, there is a relatively new field of research that seeks to make advanced physics accessible to teachers in this area, that is, physics that, due to its novelty and complexity, is beyond the reach of a typical high school teacher. For the reasons discussed in this paper, I the name “advanced physics transposition” or APT is suggested for this field of research. The objective of this work is to make a first attempt to systematise this field, to draw attention to the place it occupies, and no less important to give it a name that represents the objectives, methods and results that characterise it.\\


\begin{center}\rule{0.9\textwidth}{0.1mm} \end{center}
\end{abstract}

\maketitle

\section{Introduction}

Research into educational physics is an active field of study that has seen significant growth in recent decades. One of the most relevant areas of research is that dedicated to improving the teaching of physics in schools and colleges, where the main recipients are not professional physicists, but physics teachers in training and in practice. The leading international scientific journals in this field are Physics Education (PED), launched in 1966 [1], and The Physics Teacher (TPT), launched in 1963.\\

The fields of research in educational physics that are covered by these journals are varied, and they seek to represent all the topics that may be interesting to teachers in this area. For example, among the subjects published in PED, we find topics as diverse as new ideas on the philosophy of science which would impact on teaching; the recruitment and retention of physics teachers; initial teacher training schemes and their effectiveness; suggestions for new topics at the high school level that teachers have piloted; assessment methods; and curriculum design and content.\\

However, absent from this list and other similar ones (or, at least, poorly defined) is a relatively new field of research that seeks to make advanced physics available to teachers in the area; that is, physics that is beyond the reach of a typical high school or college teacher due to its novelty and complexity. For reasons discussed later, the name “advanced physics transposition” or APT is suggested for this field of research. Examples of advanced physics include general relativity, particle physics, quantum mechanics, and black hole physics.\\

In terms of the number of publications, APT is a comparatively small field of research, since over the last decade it has represented approximately 5\% of the total publications in educational physics. As a result, APT has not been adequately systematised, and before the present article, there was not even a proposal to give it a name that would allow it to be differentiated from other fields of study in educational physics. However, in the author's opinion, APT needs to be considered as an independent field of work within educational physics, since it has its own well-defined objectives and results; to the extent that this field of study can be differentiated and individualised, it can attract the attention of a greater number of researchers to make it grow.\\

The objective of this work is to make a first attempt to systematise this field of educational research, to draw attention to the place it occupies, and last but not least, to give it a name that covers the objectives, methods and results that characterise it. Above all, this work seeks to generate an academic dialogue between researchers who work in this field.

\section{Advanced Physics Transposition}

In the first place, we could define advanced science as that which includes active fields of research; in other words, it is that part of science where there are open problems whose clarification would represent an advance in our understanding of reality. According to another quite common meaning, we can also define advanced science as that which is characterised by a high degree of complexity and specialisation. Bringing together both meanings, we can define advanced science as that which is the subject of active research and which is usually highly complex and specialised.\\

\begin{figure}[h]
  \centering
    \includegraphics[width=0.6\textwidth]{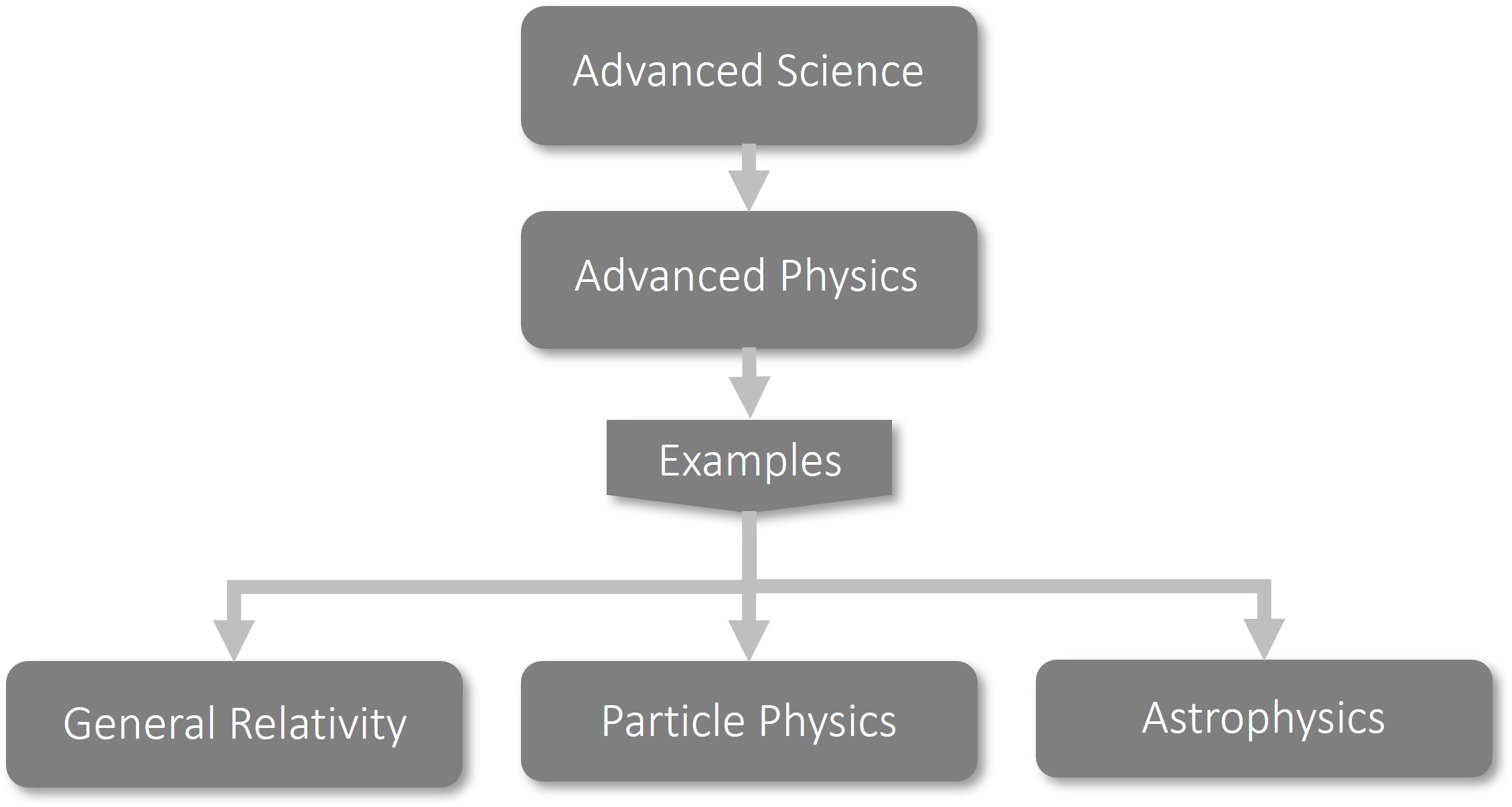}
  \caption{Relationship between advanced science and advanced physics, with some examples.}
\end{figure}

Within the area of advanced science is advanced physics, which is our main focus of interest here. In what follows, we will understand advanced physics as that part of this science that encompasses the fields of research in theoretical and experimental physics, including astronomy, which are active, and which are usually highly complex and specialised. Another way to define advanced physics is as that part of physics and astronomy that is not closed, but rather raises open questions that have not yet been addressed or agreed upon by the community of experts (Fig. 1).\\

According to our definition, advanced physics is not always recently developed knowledge. There are areas of physics that were developed more than a century ago that continue to produce advances in our understanding of reality, either because they constitute the theoretical framework of frontier research, or because their predictions or applications have not been fully explored, so that open questions and problems remain to be resolved. A good example is general relativity, which is the theory of gravity proposed by Albert Einstein [1]. Although more than a century has passed since it was proposed, Einstein's theory continues to be an active field of both theoretical and experimental research, and also constitutes the conceptual framework for a large amount of cutting-edge research [2,3].\\

\begin{figure}[h]
  \centering
    \includegraphics[width=0.6\textwidth]{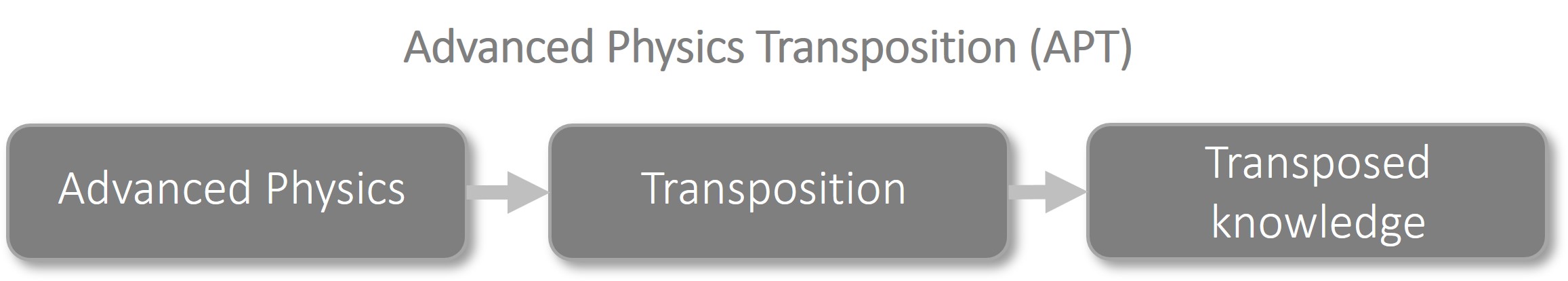}
  \caption{In general, advanced physics transposition consists of three steps: a body of knowledge to be transposed (advanced physics), the transposition process, and the transposed knowledge.}
\end{figure}

Other examples of advanced physics include quantum mechanics, particle physics, cosmology, black holes, gravitational waves, exoplanets, dark matter and dark energy, and quantum gravity. All of these research areas are very active, spark strong debate within the expert community, and are highly complex and specialised, making it very difficult for a typical physics teacher to access them directly. Examples of topics that do not fall within advanced physics include universal gravitation, classical thermodynamics, Lagrangian mechanics, and electromagnetism; although all of these fields can reach a very high degree of complexity and sophistication, they are areas of physics that from a research point of view are inactive, and are unlikely to be activated again. These areas of physics contribute to expand the frontiers of knowledge, but they do it indirectly, through its scientific or technological applications. For example, universal gravitation is used to put research satellites into orbit, and Lagrangian mechanics has recently been used to position the James Webb Space Telescope at the second Lagrange point (L2), which lies approximately 1.5 million kilometers from Earth, thereby enabling astonishing new astronomical phenomena to be discovered. However, in none of these or similar examples have universal gravitation or Lagrangian mechanics made new predictions, raises new questions or revealed new physical phenomena directly. \\

Traditionally, advanced science in general, and advanced physics in particular, have been the preserve of an elite. However, there are two main and complementary reasons why a physics teacher might be interested in advanced physics. The first and most obvious is related to personal and professional development: teachers may be interested in deepening and expanding their knowledge, and desire to go beyond the topics that are usually addressed in class. The second reason is practical: to transmit to students some of the most innovative and intriguing results in contemporary physics. In this case, the teacher is not the main recipient of advanced physics, but instead is a facilitator or a communication channel through which advanced physics reaches the students. In this scenario, advanced physics can become a complement to the school physics curriculum, the content of a non-standard course for gifted students, or the topic of a guided research project for students to go beyond traditional content, for example. In summary, we can say that a knowledge of advanced physics by teachers in this area is important because it is a source of satisfaction and personal growth, and because it is a powerful tool to enrich and complement the contents of the traditional physics curriculum, and because it can motivate students and encourage scientific vocations.

\begin{figure}[h]
  \centering
    \includegraphics[width=0.4\textwidth]{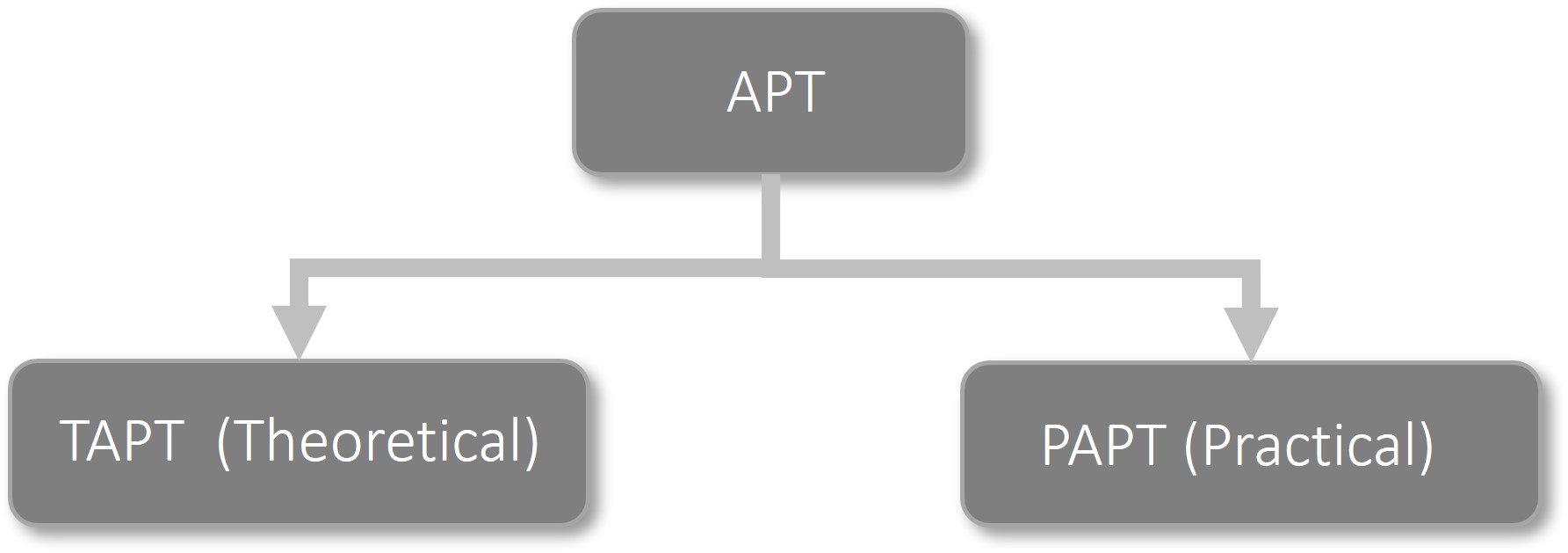}
  \caption{Research into APT may be theoretical or practical.}
\end{figure}

As noted above, due to the complexity and innovative nature of advanced physics, a typical physics teacher cannot access it directly. First, advanced physics must undergo a simplification process that brings it closer to teachers, which requires a line of research dedicated to this task. As indicated in the introduction, this line of research exists and has been developed internationally for several decades, although in an inorganic way, and without the academics who work in this field being able to clearly identify it and differentiate it from other areas of research in educational physics.\\

The first step in organising and systematising a field of study is to give it an appropriate name: I propose the name “Advanced Physics Transposition” (APT). Let us remember that advanced physics is characterised by being an active and usually complex field of research, and that the word “transposition” literally means “to put something beyond, in a different place than it originally occupied.” The term “Advanced Physics Transposition” therefore concisely describes the distinctive scope of this research area, as illustrated in Fig. 2. In summary:

\begin{shaded}
\textit{APT is a line of educational research that seeks to make advanced physics available to school and college teachers and, through them, to their students}.
\end{shaded}

Advanced physics is elitist, and APT seeks to transpose it; in other words, it seeks to bring it closer to physics teachers, and through them, to bring it closer to children and young people in schools. This transposition can be carried out both in the theoretical and practical fields. We can coin the terms TAPT for the transposition that is developed in the theoretical field and PAPT for that developed in the practical field. In general, these two types of APT research cover the same topics and only differ in terms of the approach adopted, so they are complementary. For example, in both cases, the literature contains works related to relativistic physics, particle physics, black holes, gravitational waves, etc. (Fig. 3).\\

In the case of TAPT, transposition is typically performed using simplified mathematical calculations, which seek to retain the underlying physical concepts of the detailed calculations. To do this, tools such as dimensional analysis and heuristic calculations are used. Dimensional analysis is a simple, transparent and intuitive method of obtaining approximate solutions to physics problems [4]. The main applications of dimensional analysis are the derivation of physical equations, and the verification that equations are dimensionally correct [5]. Heuristic calculations are approximate calculations based on proportionality relationships between the variables involved [6,7]. It is sometimes possible to perform the transposition without using dimensional analysis or heuristic methods; that is, it is occasionally possible to use exact mathematical treatments, but this is only viable when advanced physics can be analysed in very simple and/or highly idealised physical scenarios. An interesting example of TAPT involves articles on black holes, which describe their main characteristics and properties using mathematical tools that generally do not go beyond high school algebra [8–12]. Another example of TAPT is astrophysics, where topics as varied as dark matter, exoplanets, Hubble's law or the Chandrasekhar limit, among others, are addressed [13–20]. TAPT works rarely use calculus, and when they do, the mathematical treatment is usually simple and does not go beyond the rudiments of differentiation and integration.\\

In the case of PAPT, transposition can be carried out through two paths that are not exclusive but complementary, although these are usually presented separately. One way is to devise simple, low-cost experiments, both demonstrative and non-demonstrative, to be carried out in schools and colleges. The other is through practical activities where students can manipulate and study objects that seek to exemplify or simulate a physical phenomenon. In this case, no measurements are made or data collected, but through models and analogies, students can grasp the essential aspects of the physical phenomena of interest. One example is the use of Lego pieces or toy blocks to construct three-dimensional maps of isotopes [21], or to represent atomic masses [22]. Another interesting example is the use of dance to bring the complex concepts of particle physics closer to small children [23].\\

Although we have restricted this discussion to the needs of physics teachers, who are the main recipients of APT, it is important to keep in mind that the results of this field of research can also be beneficial for other types of education professionals, such as primary school science teachers and secondary school mathematics and science teachers.

\section{APT and didactic transposition (DT)}
APT has some similarities with DT, a term introduced by Yves Chevallard, which refers to the process by which scientific content is modified to adapt it to its teaching, transforming scholarly knowledge into teachable knowledge, appropriate to the level of students in the school stage [24,25].\\ 

Although DT can be a good complement to APT, there are at least three aspects that differentiate them. The first and most obvious is that DT covers all natural sciences, while APT is focused exclusively on physics. The second is that APT is only interested in advanced scientific knowledge, while DT deals with all scientific knowledge. The third difference is that APT does not use the theoretical framework and concepts of didactics of sciences, which gives it greater breadth and flexibility. Remember that the expression “didactics of sciences” is commonly used to describe the science of teaching science.\\

Apart from these three general differences, there is a fourth, more specific difference that mainly concerns TAPT. As we know, unlike DT, TAPT does not seek to convert scholarly knowledge into teachable knowledge; rather, it seeks to convert highly complex scholarly knowledge into less complex scholarly knowledge, intended for people with a good mathematical background, such as physics teachers.\\

This last point provides a good example of how DT can be complemented by APT. Indeed, due to their complexity, APT products are sometimes not suitable for direct introduction into the classroom, so it is the teacher's job to adapt these products to the requirements of their students. And the ideal tool to make this adaptation is DT. The reader interested in further exploring DT can consult the extensive literature on the subject. An excellent exposition can be found in  [26].

\begin{table}[htbp] 
\begin{center}
\caption{Main areas of research in physics addressed in APT}
\resizebox{1\textwidth}{!} {
\begin{tabular}{ m{6cm} m{10cm} }
\toprule
\textbf{Physics research area} & \textbf{Description} \\
\midrule

General relativity & The study of gravity, understood from a geometric point of view as a manifestation of the curvature of the space-time continuum \\ 

Special relativity & The study of the movement of bodies in the absence of gravity, based on the conception of space and time as a single entity called space-time \\ 

Astrophysics and space sciences & An interdisciplinary science in which the principles of physics are applied to the study of astronomical objects and phenomena, including cosmology, which relates to the origin, evolution and destiny of the universe as a whole  \\ 

Black hole physics & The study of the most extreme prediction of general relativity, black holes, using other areas of physics such as quantum mechanics and thermodynamics \\ 

Quantum mechanics & The study of nature at small spatial scales, atomic and subatomic systems, and their interactions. \\ 

Nuclear and particle physics  & The study of the fundamental structure of matter and the interactions between subatomic particles  \\ 
\midrule
\end{tabular}
}
\label{Main areas of research in physics addressed in APT}
\end{center}
\end{table}

\section{Advanced Physics Transposition: A Closer Look}

It is interesting to study in more detail how APT has evolved over time, and the specific topics it has addressed. To do this, we can analyse the research in educational physics conducted during the last decade, taking as a reference the two most important journals on the subject: PED and TPT. When carrying out this analysis, only manuscripts falling into the category of papers or their equivalent were considered; in other words, letters to the editor, comments, replies, etc. were not included.\\

The results of the analysis, which are presented below, required APT to be separated from other research in educational physics, which involved introducing classification criteria that, by their nature, always contain a margin of arbitrariness.\\ 

\begin{figure}[h]
  \centering
    \includegraphics[width=0.8\textwidth]{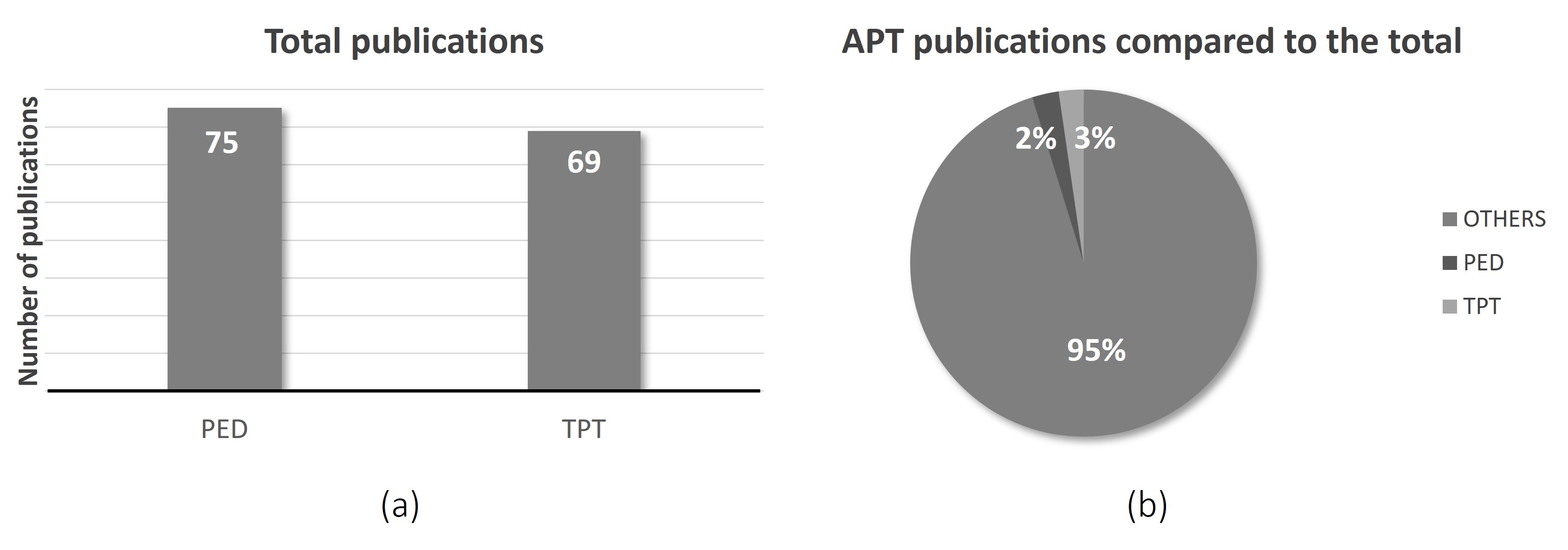}
  \caption{(a) Total publications on APT in PED and TPT in the decade 2014-2023; (b) percentages of publications on APT with respect to the total published between 2014 and 2023.}
\end{figure}

After a general review of the literature in educational physics, six major areas of physics could be identified as the objects of interest of those who work in APT. These six areas and their corresponding descriptions appear in Table 1, as follows: general relativity, special relativity, astrophysics and space sciences, black holes, quantum mechanics, nuclear and particle physics. \\ 

\begin{figure}[h]
  \centering
    \includegraphics[width=0.5\textwidth]{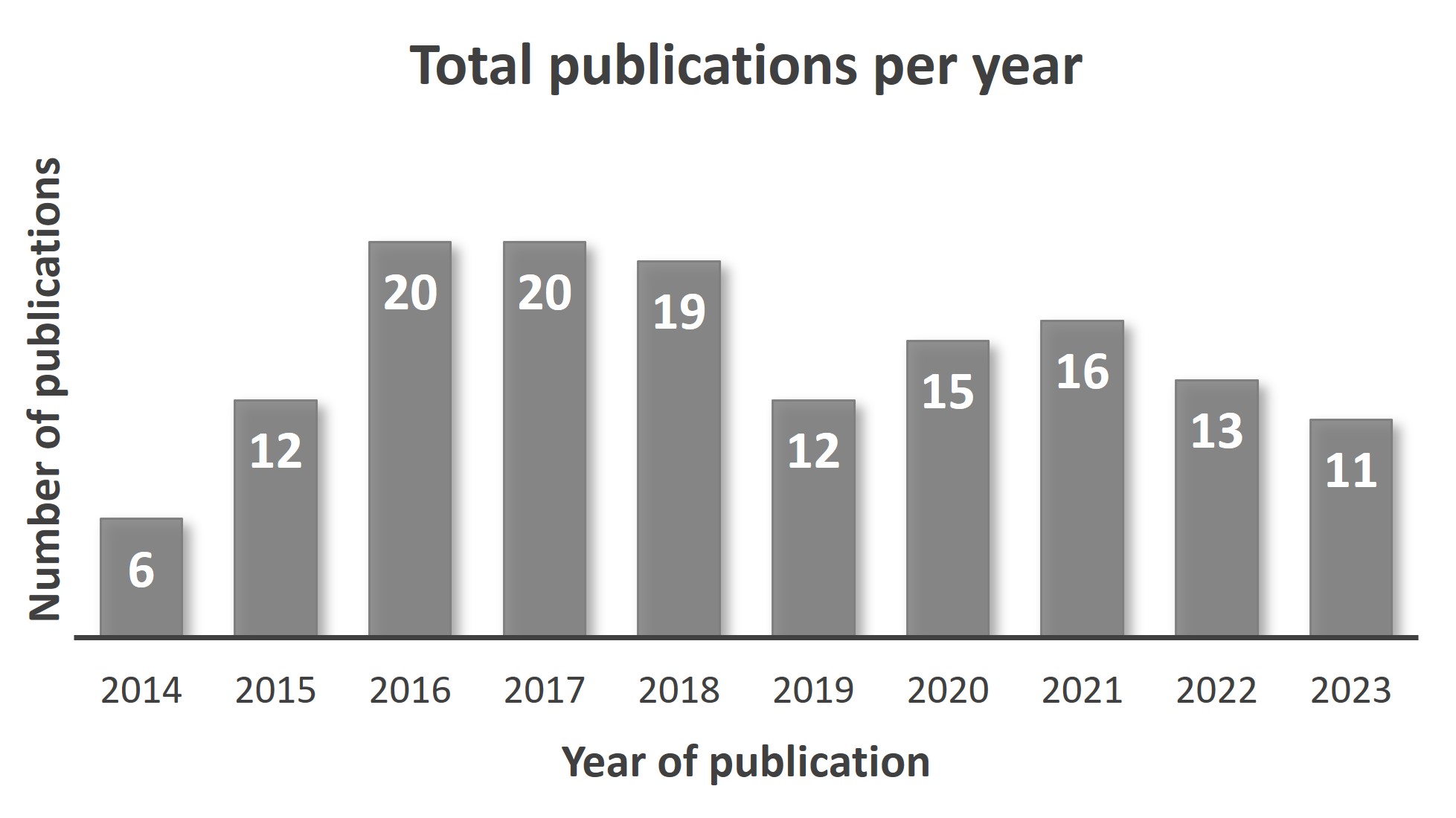}
  \caption{Total numbers of articles on APT published in PED and TPT between 2014 and 2023, separated by year.}
\end{figure}

These six areas correspond to what we have defined as advanced physics, and satisfy the definition of APT given above. These areas closely resemble the typical syllabus of a modern physics course or text, which should not be surprising, since advanced physics is also modern in a historical sense. The classification criteria used to generate the six categories in Table 1 are not explored here, since they are entirely conventional; obviously, a detailed analysis would need to make distinctions within each of these areas (astrophysics, for example, contains a large number of subfields of research), but for our purposes this is not relevant.\\ 

The graph in Fig. 4(a) shows an overview of the total number of publications on APT published in PED and TPT during the decade 2014–2023. If we add the number of publications from both journals, we find a total of 144 publications between 2014 and 2023. Fig. 4(b) shows the percentages of articles on APT compared to the total number published by PED and TPT over the last decade. It can be seen that APT represents approximately 5\% of the total publications.\\ 

\begin{figure}[h]
  \centering
    \includegraphics[width=0.8\textwidth]{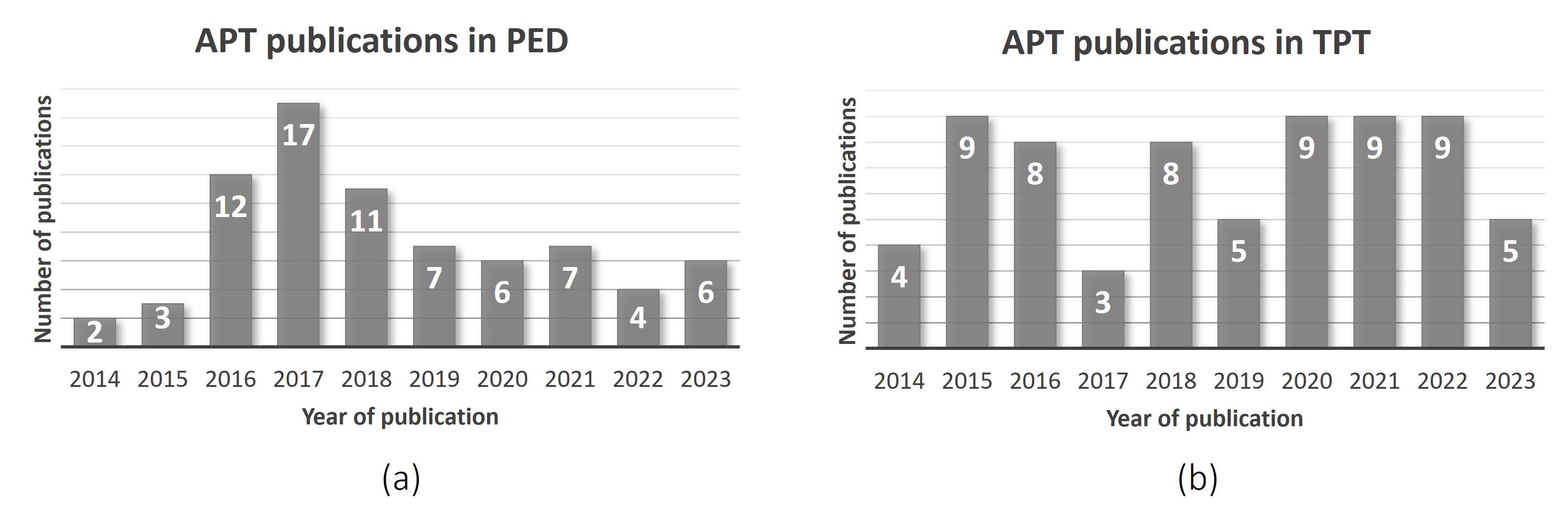}
  \caption{(a) Total articles on APT published in (a) PED and (b) TPT over the decade 2014–2023.}
\end{figure}

The graph in Fig. 5 shows the total publications during the decade 2014–2023, separated by year, while Figs. 6(a) and 6(b) show the same information, but separated by journal. The graphs do not reveal a pattern or trend, but rather random distributions of the numbers of articles.\\ 

\begin{figure}[h]
  \centering
    \includegraphics[width=0.8\textwidth]{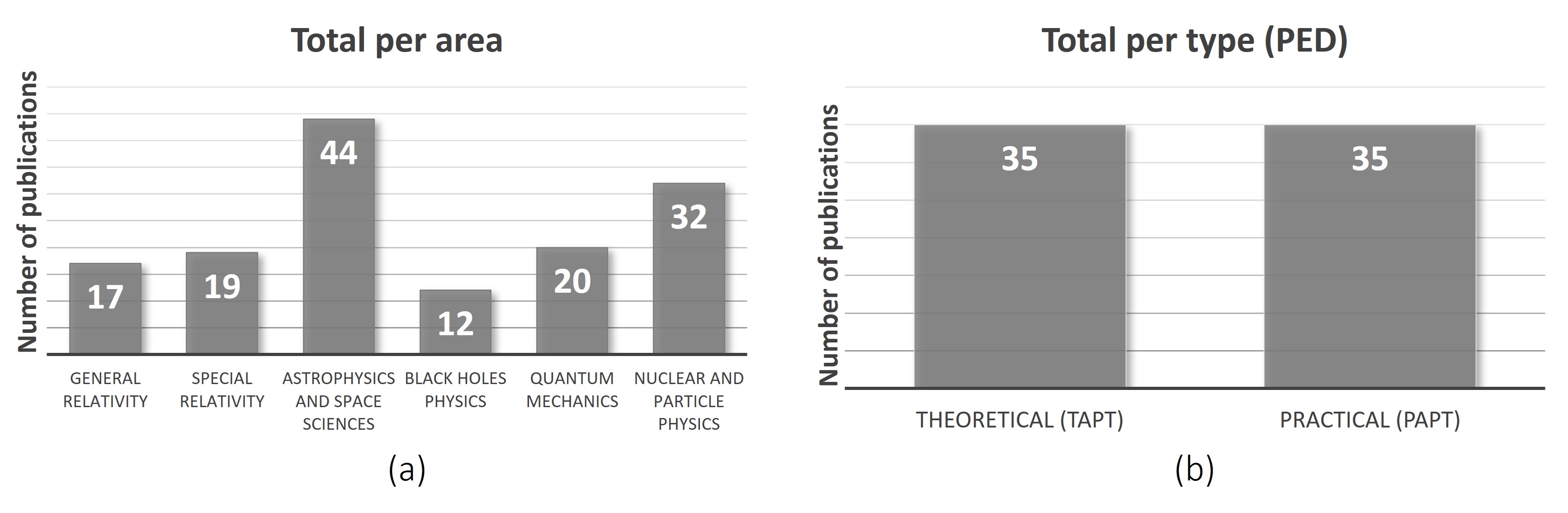}
  \caption{(a) Total publications on APT in PED and TPT separated by area; (b) total publications on APT in PED, separated by type of manuscript.}
\end{figure}

Fig. 7(a) shows the total number of manuscripts on APT published in TPT and PED in the period 2014–2023, separated according to the six major areas of physics in Table 1. It can be seen that the area in which the largest number of publications are concentrated is astrophysics and space sciences, followed by nuclear and particle physics. Finally, Fig. 7(b) shows the total publications in PED over the period 2014–2023, separated by the type of work, i.e. theoretical (TAPT) or practical (PAPT). Since TPT does not use abstracts, it was not possible to access information about the types of publication for this journal, so Fig. 7(b) shows only the data for PED. It can be observed that the total number of works is divided into equal parts. 

\section{Final Comments}

If researchers working in APT are asked to clearly indicate what they do, it is likely that they will have difficulty clearly differentiating their area of work from the multiple other areas of research in educational physics. I have found myself in this situation many times, and this has formed one of my main motivations for writing this manuscript.\\

As a brief synthesis of the ideas developed here, we can say that APT is a line of educational research that seeks to make advanced physics available to school and college teachers, where “advanced physics” is understood as that part of the science that includes active research topics, both theoretical and experimental, which are frequently highly complex and specialised. In other words, the overall objective of APT is to transpose advanced physics, which means bringing it closer to teachers and, through them, to their students.\\

It is common for APT researchers to have gained PhDs in physics or astronomy, and in parallel with their work in their respective fields, to make forays into educational physics in order to transpose frontier topics. However, school and college physics teachers also work actively in APT. This shows us that APT requires a rather specific type of educational researcher, who, together with a deep and up-to-date knowledge of physics and a solid experimental training, has a clear awareness of the needs of teachers in this area. In short, this is a type of researcher who has the rare ability to make difficult things easy.\\

It is hoped that this manuscript has fulfilled its main objectives, which include highlighting the convenience of considering APT as an independent research area within educational physics, and identifying its main objectives and results. Above all, it is hoped that this manuscript will generate a constructive dialogue among the researchers who work in this field, which is already sufficiently mature to become independent and come to light.

\section*{Acknowledgments}
I would like to thank to Daniela Balieiro for their valuable comments in the writing of this paper. 

\section*{References}

[1] G. Williams, Happy Birthday Physics Education, Phys. Educ. 51 (2016) 030101. https://doi.org/10.1088/0031-9120/51/3/030101.

\vspace{2mm}

[2] A. Einstein, Die Grundlage der allgemeinen Relativitätstheorie, Annalen Der Physik 354 (1916) 769–822.

\vspace{2mm}

[3] J. Pinochet, Classical Tests of General Relativity Part I: Looking to the Past to Understand the Present, Physics Education 55 (2020) 65016.

\vspace{2mm}

[4] J. Pinochet, General relativity in a nutshell I, Phys. Scr. 98 (2023) 126103. https://doi.org/10.1088/1402-4896/ad0c34.

\vspace{2mm}

[5] D.S. Lemons, A Student’s Guide to Dimensional Analysis, Cambridge University Press, Cambridge, 2017.

\vspace{2mm}

[6] E.F. Redish, Using Math in Physics: 1. Dimensional Analysis, The Physics Teacher 59 (2021) 397–400. https://doi.org/10.1119/5.0021244.

\vspace{2mm}

[7] S. Mahajan, The Art of Insight in Science and Engineering, The MIT Press, Cambridge, 2014.

\vspace{2mm}

[8] S. Mahajan, Street-Fighting Mathematics, The MIT Press, Cambridge, 2020.

\vspace{2mm}

[9] I.A. Barr, A. Bull, E. O’Brien, K.A. Drillsma-Milgrom, L.R. Milgrom, Illuminating black holes, Phys. Educ. 51 (2016) 043001. https://doi.org/10.1088/0031-9120/51/4/043001.

\vspace{2mm}

[10] S.V. Kontomaris, A. Malamou, A presentation of the black hole stretching effect, Phys. Educ. 53 (2017) 015010. https://doi.org/10.1088/1361-6552/aa8d22.

\vspace{2mm}

[11] J. Pinochet, “Black holes ain’t so black”: An introduction to the great discoveries of Stephen Hawking, Phys. Educ. 54 (2019) 035014.

\vspace{2mm}

[12] J. Pinochet, Five misconceptions about black holes, Phys. Educ. 54 (2019) 55003.
[13] J. Pinochet, The little robot, black holes, and spaghettification, Phys. Educ. 57 (2022) 045008. https://doi.org/10.1088/1361-6552/ac5727.

\vspace{2mm}

[14] M. Cowley, S. Hughes, Characterization of transiting exoplanets by way of differential photometry, Phys. Educ. 49 (2014) 293. https://doi.org/10.1088/0031-9120/49/3/293.

\vspace{2mm}

[15] D. Della-Rose, R. Carlson, K. de La Harpe, S. Novotny, D. Polsgrove, Exoplanet Science in the Classroom: Learning Activities for an Introductory Physics Course, The Physics Teacher 56 (2018) 170–173. https://doi.org/10.1119/1.5025299.

\vspace{2mm}

[16] J. Ford, J. Stang, C. Anderson, Simulating Gravity: Dark Matter and Gravitational Lensing in the Classroom, The Physics Teacher 53 (2015) 557–560. https://doi.org/10.1119/1.4935771.

\vspace{2mm}

[17] J.M. Hyde, Exploring Hubble Constant Data in an Introductory Course, The Physics Teacher 59 (2021) 159–161. https://doi.org/10.1119/10.0003654.

\vspace{2mm}

[18] A.M. Low, The Chandrasekhar limit: a simplified approach, Phys. Educ. 58 (2023) 045008. https://doi.org/10.1088/1361-6552/acdbb0.

\vspace{2mm}

[19] Y.-Z. Ma, S.-N. Zhang, Hubble expansion is not a velocity, Phys. Educ. 51 (2016) 065011. https://doi.org/10.1088/0031-9120/51/6/065011.

\vspace{2mm}

[20] J. Pinochet, Illuminating dark matter: II. A guide for physics teachers, Phys. Educ. 59 (2024) 045002. https://doi.org/10.1088/1361-6552/ad3605.

\vspace{2mm}

[21] C.A. Diget, A. Pastore, K. Leech, T. Haylett, S. Lock, T. Sanders, M. Shelley, H.V. Willett, J. Keegans, L. Sinclair, E.C. Simpson,  the B.B. Collaboration, Binding blocks: building the Universe one nucleus at a time, Phys. Educ. 52 (2017) 024001. https://doi.org/10.1088/1361-6552/aa550c.

\vspace{2mm}

[22] H. Koura, Three-dimensional nuclear chart—understanding nuclear physics and nucleosynthesis in stars, Phys. Educ. 49 (2014) 215. https://doi.org/10.1088/0031-9120/49/2/215.

\vspace{2mm}

[23] K. Nikolopoulos, M. Pardalaki, Particle dance: particle physics in the dance studio, Phys. Educ. 55 (2020) 025018. https://doi.org/10.1088/1361-6552/ab6952.

\vspace{2mm}

[24] Y. Chevallard, La transposition didactique: Du savoir savant au savoir enseigné [Didactic transposition: From scholarly knowledge to taught knowledge]. La Pensée Sauvage Editions, Grenoble, (1991).

\vspace{2mm}

[25] Y. Chevallard, M. Bosch, Didactic Transposition in Mathematics Education. In: Lerman, S. (eds) Encyclopedia of Mathematics Education. Springer, Dordrecht, (2014). https://doi.org/10.1007/978-94-007-4978-8\_48.

\vspace{2mm}

[26] M. Achiam, Didactic transposition: From theoretical notion to research programme. Paper presented at the biannual ESERA (European Science Education Research Association) doctoral summer school August 25-29 in Kappadokya, Turkey, (2014).

\end{document}